\documentclass[aps,11pt,pra,showpacs,superscriptaddress,preprint]{revtex4}

\usepackage{amsmath}

\usepackage{graphicx}

\usepackage{subfig}

\usepackage{array}

\catcode`ð=\active

\defð{\u{g}}

\catcode`Ð=\active

\defÐ{\u{G}}

\catcode`Ý=\active

\defÝ{\. I}

\catcode`ö=\active

\defö{\"{o}}

\catcode`Ö=\active

\defÖ{\"O}

\catcode`ü=\active

\defü{\"{u}}

\catcode`Ü=\active

\defÜ{\"{U}}

\catcode`Þ=\active

\defÞ{\c{S}}

\catcode`þ=\active

\defþ{\c{s}}

\catcode`ý=\active

\defý{{\i}}

\catcode`ç=\active

\defç{\d{c}}

\catcode`Ç=\active

\defÇ{\d{C}}

\begin{document}
\title{Approximate analytical solutions of the Dirac equation for Yukawa
potential plus Tensor Interaction with any $\kappa$-value}
\author{\small Altuð Arda}
\email[E-mail: ]{arda@hacettepe.edu.tr}\affiliation{Department of
Physics Education, Hacettepe University, 06800, Ankara,Turkey}
\author{\small Ramazan Sever}
\email[E-mail: ]{sever@metu.edu.tr}\affiliation{Department of
Physics, Middle East Technical  University, 06531, Ankara,Turkey}

\begin{abstract}
Approximate analytical solutions of the Dirac equation are
obtained for the Yukawa potential plus a tensor interaction with
any $\kappa$-value for the cases having the Dirac equation
pseudospin and spin symmetry. The potential describing tensor
interaction has a Yukawa-like form. Closed forms of the energy
eigenvalue equations and the spinor wave functions are computed by
using the Nikiforov-Uvarov method. It is observed that the energy
eigenvalue equations are consistent with the ones obtained before.
Our numerical results are also listed to see the
effect of the tensor interaction on the bound states.

\end{abstract}


\maketitle

\newpage

\section{Introduction}

The Yukawa potential has the form [1]
\begin{eqnarray}
V(r)=-\frac{\eta}{r}\,e^{-\beta r}\,.
\end{eqnarray}
Here $\beta$ is the screening parameter and $\eta$ is the
potential strength. This potential is one of the different types
of the screened Coulomb potentials which have been studied in
various areas of physics such as atomic physics, plasma physics,
solid-state and astrophysics [2]. The readers can find an outline
about the approaches used to study this potential in Ref. [3]. The
Yukawa potential is a basic ground to describe interaction between
charged particles in colloidal suspensions [4] and an important
potential model in Thomas-Fermi approximation to the electron gas
[5]. In the view of the present work, it should be noted that,
with its other properties, the Yukawa potential has bound states
only for the values of the parameter $\beta$ below a value
$\beta_{c}<1.19$ (in a.u.) [6]. The Yukawa potential [7-15] has
also received a great deal of attention in view of the methods by
which the potential has been studied. Some of them used to solving
the wave equations are the $1/N$-expansion [16] and shifted
$1/N$-expansion methods [17], studying the potential by using the
Raygleih-Schrödinger perturbation expansion [18], a
group-theoretical approach by using the Fock transformation [3],
the variational self-consistent field molecular-orbital method
[18], the J-matrix method [19], a new approximation scheme
proposed to study the bound states of potential [20], studying in
terms of the hypervirial theorems [21] and a numerical solution of
the Schrödinger equation for the present potential [22].

In the present work, we study the problem including also the
two-component spinor wave functions in terms of the hypergeometric
functions within the context of pseudospin and spin symmetries
[23-25]. The Dirac equation with vector, $V(r)$, and scalar,
$S(r)$, potentials has pseudospin (spin) symmetry when the
difference (the sum) of the potentials $V(r)-S(r)$ $[V(r)+S(r)]$
is constant, which means $\frac{d}{dr}[V(r)-S(r)]=0$ (or
$\frac{d}{dr}[V(r)+S(r)]=0$). It is pointed out that these
symmetries can explain degeneracies in single-particle energy
levels in nuclei or in some heavy meson-spectra within the context
of relativistic mean-field theories [23-35]. In the relativistic
domain, these symmetries were used in the context of deformation
and superdeformation in nuclei, magnetic moment interpretation and
identical bands [26]. In the non-relativistic domain, performing a
helicity unitary transformation to a single-particle Hamiltonian
maps the normal state onto the pseudo-state [27]. Because of these
investigations, the solutions of the Dirac equation having spin
and pseudospin symmetry have received great attention for
different type of potentials such as Morse potential, Eckart
potential, etc. [28-32].

In Ref. [33] the pseudospin symmetric solutions of the Dirac
equation are obtained for the harmonic oscillator while used the
tensor interaction as a potential linear in $r$. In Ref. [34] spin
and pseudospin symmetric solutions are studied for the Woods-Saxon
potential by taking the tensor interaction as a Coulomb-like
potential. In this point of view, we propose to use a Yukawa-like
potential as the tensor interaction
\begin{eqnarray}
U(r)=\,\nu\frac{e^{-\beta r}}{r}\,,
\end{eqnarray}
which has an attractive form. Choosing this form makes it possible
to find analytical solutions of the present problem.

The organization of this work is as follows. In Section 2, we
briefly give the Dirac equation with attractive scalar and
repulsive vector potentials for the cases where the Dirac equation
has pseudospin and spin symmetry, respectively. In Section 3, we
present the Nikiforov-Uvarov (NU) method and the parameters
required within the method. In Section 4, we find an analytical
expression for the bound states and the two-component spinor wave
functions of the Yukawa potential by using an approximation
instead of the spin-orbit coupling term. We analyze the problem
for the cases having the Dirac equation pseudospin and also spin
symmetry. We give the numerical energy eigenvalues for the
different quantum number pairs $(n,\kappa)$ where we choose the
parameter and the mass values in $a.u.$. The last section includes
our conclusions.

\section{Dirac Equation and Spin and Pseudospin Symmetry}

The Dirac equation is basically written by using linear momentum
operator, $P_{\mu}=i\hbar\partial_{\mu}$ (four-vector) and the
scalar rest mass $M$. As a result, two potential couplings are
used in equation. One coupling is a gauge invariant one to the
four-vector potential $A_{\mu}(\vec{r},t)$ by using $P_{\mu}
\rightarrow P_{\mu}-gA_{\mu}$ ($g$ is a real coupling parameter)
and the other one is to the space-time scalar potential
$S(\vec{r},t)$ by substitution $M \rightarrow M+S$. The
"four-vector" and "scalar" terms mean that classifying the
observable according to the unitary irreducible representation of
the rotation and translation groups in Minkowski space-time. By
taking the space component of the vector potential to vanish
($\vec{A}=0$) and writing the time component of the four-vector
potential as $gA_{0}=V(\vec{r},t)$, then we obtain the so-called
vector and scalar potentials, $V(r)$ and $S(r)$, respectively.

The free particle Dirac equation is given ($\hbar=c=1$)
\begin{eqnarray}
\bigl(i\gamma^{\mu}\partial_{\mu}-M\bigr)\Psi(\vec{r},t)=0\,,
\end{eqnarray}
and taking the total wave function as
$\Psi(\vec{r},t)=e^{-iEt}\psi(\vec{r})$ for time-independent
potentials, where $E$ is the relativistic energy, the above
equation including also a tensor interaction, $U(r)$, with
spherical symmetric vector and scalar potentials is written as
\begin{eqnarray}
\bigl[\vec{\alpha}.\vec{P}+\beta
M-i\beta\vec{\alpha}.\hat{r}U(r)+\beta
S(r)\bigr]\psi(\vec{r})=\big[E-V(r)\bigr]\psi(\vec{r})\,,
\end{eqnarray}
Here $\alpha$ and $\beta$ are usual $4\times4$ matrices. For
spherical nuclei, the angular momentum $\vec{J}$ and the operator
$\hat{K}=-\beta\bigl(\hat{\sigma}.\hat{L}+1\bigr)$ with
eigenvalues $\kappa=\pm (j+1/2)$ commute with the Dirac
Hamiltonian, where $\hat{L}$ is the orbital angular momentum. By
using the radial eigenfunctions for upper and lower components of
the Dirac eigenfunction $F(r)$ and $G(r)$, respectively, the wave
function is written as [35]
\begin{eqnarray}
\psi(\vec{r})=\,\frac{1}{r}\,\Bigg[\begin{array}{c}
\,F\,(r)Y^{(1)}(\theta,\phi) \\
iG\,(r)Y^{(2)}(\theta,\phi)
\end{array}\Bigg]\,,
\end{eqnarray}
where $Y^{(1)}(\theta,\phi)$ and $Y^{(2)}(\theta,\phi)$ are the
pseudospin and spin spherical harmonics, respectively. They
correspond to angular and spin parts of the wave function given by
\begin{eqnarray}
Y^{(1),(2)}(\theta,\phi)=\sum_{m_{\ell}m_{s}}<\ell
m_{\ell}\frac{1}{2}m_{s}|\ell\frac{1}{2}jm>Y_{\ell
m_{\ell}}(\theta,\phi)\chi_{\frac{1}{2}m_{s}}\,,\nonumber\\j=|\kappa|-\frac{1}{2}\,,\,\,\,
\ell=\kappa\,\,(\kappa>0)\,;\,\ell=-(\kappa+1)\,\,(\kappa<0)\,,
\end{eqnarray}
Here, $Y_{\ell m_{\ell}}(\theta,\phi)$ denotes the spherical
harmonics and $m_{\ell}$ and $m_{s}$ are related magnetic quantum
numbers.

Substituting Eq. (5) into Eq. (4) and using the followings
\begin{subequations}
\begin{align}
\bigl(\vec{\sigma}.\vec{L}\bigr)Y^{(2)}(\theta,\phi)&=(\kappa-1)Y^{(2)}(\theta,\phi)\,,\\
\bigl(\vec{\sigma}.\vec{L}\bigr)Y^{(1)}(\theta,\phi)&=-(\kappa-1)Y^{(1)}(\theta,\phi)\,,\\
\bigl(\vec{\sigma}.\hat{r}\bigr)Y^{(2)}(\theta,\phi)&=-Y^{(1)}(\theta,\phi)\,,\\
\bigl(\vec{\sigma}.\hat{r}\bigr)Y^{(1)}(\theta,\phi)&=-Y^{(2)}(\theta,\phi)\,,
\end{align}
\end{subequations}
give us the following coupled differential equations
\begin{subequations}
\begin{align}
&&\left(\frac{d}{dr}+\frac{\kappa}{r}-U(r)\right)F(r)=[E+M-\Gamma(r)]G(r)\,,\\
&&\left(\frac{d}{dr}-\frac{\kappa}{r}+U(r)\right)G(r)=[M-E+\Lambda(r)]F(r)\,.
\end{align}
\end{subequations}
where $\Gamma(r)=V(r)-S(r)$ and $\Lambda(r)=V(r)+S(r)$. Using the
expression $G(r)$ in Eq. (8a) and inserting it into Eq. (8b), we
get a second order differential equation
\begin{eqnarray}
\left[\frac{d^2}{dr^2}-\frac{\kappa(\kappa+1)}{r^2}+\varepsilon^{(1)}(r)+
\bigr(\frac{2\kappa}{r}-U(r)-\frac{d}{dr}\bigl)U(r)\right]F(r)
=-\left[\frac{d\Gamma(r)/dr}{\left[E+M-\Gamma(r)\right]}\right]F(r)\,,
\end{eqnarray}
where $\varepsilon^{(1)}(r)=\left[E+M-\Gamma(r)\right]
\left[E-M-\Lambda(r)\right]$. By similar steps, we write the
following second order differential equation for $G(r)$ as
\begin{eqnarray}
\left[\frac{d^2}{dr^2}-\frac{\kappa(\kappa-1)}{r^2}+\varepsilon^{(2)}(r)+
\bigr(\frac{2\kappa}{r}-U(r)+\frac{d}{dr}\bigl)\right]
G(r)=\left[\frac{d\Lambda(r)/dr}{\left[M-E+\Lambda(r)\right]}\right]G(r)\,,
\end{eqnarray}
where $\varepsilon^{(2)}(r)=\left[E-M-\Lambda(r)\right]
\left[E+M-\Gamma(r)\right]$. The last two equations have the
following forms
\begin{subequations}
\begin{align}
&&\left\{\frac{d^2}{dr^2}-\frac{\kappa(\kappa+1)}{r^2}+\bigr(\frac{2\kappa}{r}
-U(r)-\frac{d}{dr}\bigl)U(r)+\left[E+M-A\right]
\left[E-M-\Lambda(r)\right]\right\}F(r)=0\,,\\
&&\left\{\frac{d^2}{dr^2}-\frac{\kappa(\kappa-1)}{r^2}+\bigr(\frac{2\kappa}{r}
-U(r)+\frac{d}{dr}\bigl)U(r)+\left[E-M-A\right]
\left[E+M-\Gamma(r)\right]\right\}G(r)=0\,.
\end{align}
\end{subequations}
if the Dirac equation has spin symmetry which means that
$\Gamma(r)=A$ ($d\Gamma(r)/dr=0$) is a constant and pseudospin
symmetry which means $\Lambda(r)=A$ ($d\Lambda(r)/dr=0$) is a
constant [25-27].

\section{Nikiforov Uvarov Method}

The Nikiforov-Uvarov method could be used to solve a second-order
differential equation of the hypergeometric-type which can be
transformed by using appropriate coordinate transformation into
the following form
\begin{eqnarray}
\sigma^2(t)\frac{d^2\Psi(t)}{dt^2}+\sigma(t)\tilde{\tau}(t)\frac{d\Psi(t)}{dt}+\tilde{\sigma}(t)
\Psi(t)=0\,,
\end{eqnarray}
where $\sigma(t)$\,, and $\tilde{\sigma}(t)$ are polynomials, at
most, second degree, and $\tilde{\tau}(t)$ is a first-degree
polynomial. By taking the solution as
\begin{eqnarray}
\Psi(t)=\psi(t)\varphi(t)\,,
\end{eqnarray}
gives Eq. (12) as a hypergeometric type equation [36]
\begin{eqnarray}
\frac{d^2\varphi(t)}{dt^2}+\frac{\tau(t)}{\sigma(t)}\frac{d\varphi(t)}{dt}+\frac{\lambda}{\sigma(t)}\,
\varphi(t)=0\,,
\end{eqnarray}
where $\psi(t)$ is defined by using the equation [36]
\begin{eqnarray}
\frac{1}{\psi(t)}\frac{d\psi(t)}{dt}=\frac{\pi(t)}{\sigma(t)}\,,
\end{eqnarray}
and the other part of the solution in Eq. (13) is given by
\begin{eqnarray}
\varphi_{n}(t)=\frac{a_{n}}{\rho(t)}\frac{d^n}{dz^n}[\sigma^{n}(z)\rho(t)]\,,
\end{eqnarray}
where $a_{n}$ is a normalization constant, and $\rho(t)$ is the
weight function, and satisfies the following equation [36]
\begin{eqnarray}
\frac{d\sigma(t)}{dt}+\frac{\sigma(t)}{\rho(t)}\frac{d\rho(t)}{dt}=\tau(t)\,.
\end{eqnarray}

The function $\pi(t)$ and the parameter $\lambda$ in the above
equation are defined as
\begin{eqnarray}
\pi(t)&=&\,\frac{1}{2}\,[\frac{d}{dt}\,\sigma(t)-\tilde{\tau}(t)]
\pm\bigg\{\frac{1}{4}\left[\frac{d}{dt}\,\sigma(t)-\tilde{\tau}(t)\right]^2
-\tilde{\sigma}(t)+k\sigma(t)\bigg\}^{1/2}\,,\\
\lambda&=&k+\frac{d}{dt}\,\pi(t)\,.
\end{eqnarray}

In the NU method, the square root in Eq. (18) must be the square
of a polynomial, so the parameter $k$ can be determined. Thus, a
new eigenvalue equation becomes
\begin{eqnarray}
\lambda=\lambda_{n}=-n\frac{d}{dt}\,\tau(t)-\frac{1}{2}\,(n^2-n)\frac{d^2}{dt^2}\,\sigma(t)\,.
\end{eqnarray}
where prime denotes the derivative, and the derivative of the
function $\tau(t)=\tilde{\tau}(t)+2\pi(t)$ should be negative.

\section{Bound State Solutions}

\subsubsection{Pseudospin Symmetric Case}
Taking the scalar and vector potentials in Eq. (4) as
\begin{eqnarray}
S(r)=-\frac{\eta_{s}}{r}\,e^{-\beta
r}\,\,;\,\,\,V(r)=+\frac{\eta_{v}}{r}\,e^{-\beta r}\,,
\end{eqnarray}
inserting them into Eq. (11b), using Eq. (2) and taking the
following expression instead of the spin-orbit coupling term [37,
38, 39]
\begin{eqnarray}
\frac{1}{r^2}\simeq \beta^2\,\frac{1}{(1-e^{-\beta r })^2}
\end{eqnarray}
we obtain from Eq. (11b)
\begin{eqnarray}
\bigg\{\frac{d^2}{dr^2}-\frac{\beta^2\kappa(\kappa-1)}{(1-e^{-\beta
r })^2}+\bigl[\bigl(2\kappa\nu\beta^2-\nu\beta^2\bigr)e^{-\beta r
}-\nu^2\beta^2e^{-2\beta r }\bigr]\,\frac{1}{(1-e^{-\beta r
})^2}\nonumber\\-\nu\beta^2\,\frac{e^{-\beta r}}{1-e^{-\beta r }}
-\beta\eta_{1}\left[E-M-A\right]\frac{e^{-\beta r}}{1-e^{-\beta r
}}+\left[E-M-A\right]\left[E+M\right]\bigg\}G(r)=0\,,\nonumber\\
\end{eqnarray}
where $\eta_{1}=\eta_{s}+\eta_{v}$. Using a new variable
$t=1-e^{-2\beta r}$ ($0\leq t \leq 1$) and the following
abbreviations
\begin{subequations}
\begin{align}
-a^2_{1}&=-\kappa(\kappa-1)+\nu(2\kappa-\nu-1)\,,\\
-a^2_{2}&=\nu(-2\kappa+2\nu)-\frac{\eta_{1}}{\beta}\,(E-M-A)\,,\\
-a^2_{3}&=\nu(-\nu+1)+\frac{\eta_{1}}{\beta}\,(E-M-A)+\frac{1}{\beta^2}\,(E-M-A)(E+M)\,.
\end{align}
\end{subequations}
gives us
\begin{eqnarray}
\left\{\frac{d^2}{dt^2}-\frac{t}{t(1-t)}\frac{d}{dt}+\frac{1}{t^2(1-t)^2}[-a^2_{1}-a^2_{2}t-a^2_{3}t^2]\right\}G(t)=0\,,
\end{eqnarray}

Comparing Eq. (25) with Eq. (12)
\begin{eqnarray}
\tilde{\tau}(t)=-t\,,\,\,\,\,\,\sigma(t)=t(1-t)\,,\,\,\,\,\,
\tilde{\sigma}(z)=-a_3^2t^2-a_2^2t-a_1^2\,,
\end{eqnarray}
Substituting this into Eq. (18), we get
\begin{eqnarray}
\pi(t)=\,\frac{1-t}{2}\pm\sqrt{(\frac{1}{4}+a_3^2-k)t^2+(-\frac{1}{2}+a_2^2+k)t+\frac{1}{4}+a_1^2}.
\end{eqnarray}

The constant $k$ can be determined by the condition that the
discriminant of the expression under the square root has to be
zero
\begin{eqnarray}
\Delta=(-\frac{1}{2}+a_2^2+k)^2-4(\frac{1}{4}+a_1^2)(\frac{1}{4}+a_3^2-k)=0\,.
\end{eqnarray}
The roots of $k$ are $k_{1,2}=\,-a_2^2\,-2a^2_{1}\mp\,A$, where
$A=\sqrt{(a_3^2+a_2^2+a_1^2)(1+4a^2_{1})}$. Substituting these
values into Eq. (18), we obtain $\pi(t)$ for $k_1$ as
\begin{eqnarray}
\pi(t)=\frac{1-t}{2}+
\left[\left(\sqrt{\frac{1}{4}+a^2_{1}\,}+\frac{A}{2\sqrt{\frac{1}{4}+a^2_{1}\,}}\right)t
-\sqrt{\frac{1}{4}+a^2_{1}\,}\right]\,,
\end{eqnarray}
and for $k_2$ as
\begin{eqnarray}
\pi(t)=\frac{1-t}{2}-
\left[\left(\sqrt{\frac{1}{4}+a^2_{1}\,}+\frac{A}{2\sqrt{\frac{1}{4}+a^2_{1}\,}}\right)t
-\sqrt{\frac{1}{4}+a^2_{1}\,}\right]\,.
\end{eqnarray}

Now we find the polynomial $\tau(t)$ from $\pi(t)$ as

\begin{eqnarray}
\tau(t)=1+2\sqrt{\frac{1}{4}+a^2_{1}\,}-\left(2+2\sqrt{\frac{1}{4}+a^2_{1}\,}
+\frac{A}{\sqrt{\frac{1}{4}+a^2_{1}\,}}\right).
\end{eqnarray}
so its derivative is negative. We have from Eq. (19)
\begin{eqnarray}
\lambda=\,-\,a^2_{2}-2a^2_{1}-A-\frac{1}{2}-\frac{1}{2}\,\sqrt{1+4a^2_{1}\,}-\frac{A}{\sqrt{1+4a^2_{1}\,}}\,,
\end{eqnarray}
and Eq. (20) gives us
\begin{eqnarray}
\lambda_n=n\left(2+\sqrt{1+4a^2_{1}\,}+\frac{2A}{\sqrt{1+4a^2_{1}\,}}\right)+n(n-1)\,.
\end{eqnarray}

Substituting the values of the parameters given by Eq. (24) and
setting $\lambda=\lambda_n$, one can find the following
\begin{eqnarray}
2\sqrt{\kappa(\kappa-1)-\frac{1}{\beta^2}\,(E-M-A)(E+M)\,}
+2\sqrt{\nu(\nu-1)-\frac{1}{\beta}(E-M-A)\bigl(\eta_{1}+\frac{E+M}{\beta}\bigr)\,}\nonumber\\
+\sqrt{(1-2\kappa)^2+4\nu(1+\nu-2\kappa)\,}+2n+1=0\,.
\end{eqnarray}

Thus one can find energy eigenvalues for the case of pseudospin
symmetry for any $\kappa$-value in the existence of a tensor
interaction. This analytical result is consistent with the ones
obtained in Ref. [8] for the absence of the tensor interaction.
Table I presents the numerical energy eigenvalues for different
quantum number pairs ($n,\kappa$) and different values for $\nu$
which makes it possible to see the effect of the tensor
interaction on the energy eigenstates. It is seen that energy
values are negative as stated in the theory [25-27]. Our parameter
values are as follow: $\eta_{1}=2.5$ a.u., $\beta=0.5$ a.u.,
$A=0$. We observe that the energy eigenvalues decrease while the
principal quantum number increases or spin-orbit quantum number
$\kappa$ decreases.

In order to find the eigenfunctions, we first compute the weight
function from Eq. (17)
\begin{eqnarray}
\rho(t)=t^{2a_1}(1-t)^{A}\,,
\end{eqnarray}
and the wave function becomes
\begin{eqnarray}
\varphi_{n\ell}\,(z)\sim\,\frac{1}{t^{2a_1}(1-t)^{A}}\,\frac{d^n}{dt^n}\,\left[
\,t^{n+2a_1}\,(1-t)^{n+A}\right]\,.
\end{eqnarray}
The polynomial solutions can be written in terms of the Jacobi
polynomials [40]
\begin{eqnarray}
\varphi_{n\ell}\,(t)\sim P_n^{(A,\,\,
2a_1)}\,(2t-1)\,,\,\,\,\,\,A>-1\,,\,\,\,\,\,a_1>-1\,.
\end{eqnarray}
On the other hand, the other part of the wave function is obtained
from Eq. (15) as
\begin{eqnarray}
\psi(t)=t^{a_1}(1-t)^{(1+A)/2}\,.
\end{eqnarray}
Thus, the radial eigenfunctions for the lower component of the
Dirac eigenfunction take
\begin{eqnarray}
G(t)\sim t^{a_1}(1-t)^{(1+A)/2}\,P_n^{(A,\,\,2a_1)}\,(2t-1)\,,
\end{eqnarray}
and the other radial component is obtained from Eq. (8a) as
\begin{eqnarray}
&&F(t) \sim
\frac{t^{a_1}(1-t)^{(1+A)/2}}{M-E+A}\bigg\{\beta(1-t)\bigg[\left(\frac{a_{1}}{t}
-\frac{1+A}{2(1-t)}\right)P_n^{(A,\,\,2a_1)}\,(2t-1)\nonumber\\&+&(n+A+2a_{1}+1)P_{n-1}^{(A+1,\,\,2a_{1}+1)}\,(2t-1)\bigg]
\nonumber\\&-&\frac{\kappa\beta}{lnt}
P_n^{(A,\,\,2a_1)}\,(2t-1)-\frac{\beta\nu}{lnt}(1-t)P_n^{(A,\,\,2a_1)}\,(2t-1)\bigg\}\,.
\end{eqnarray}
where we have used the property of the Jacobi polynomials as
$\frac{d}{dx}[P_{n}^{(q,\,r)}(x)]=\frac{1}{2}(n+q+r+1)P_{n-1}^{(q+1,\,r+1)}(x)$
[40].

\subsubsection{Spin Symmetric Case}

In this case, from Eqs. (2), (21) and (22), Eq. (11a) takes the
form
\begin{eqnarray}
\bigg\{\frac{d^2}{dr^2}-\frac{\beta^2\kappa(\kappa-1)}{(1-e^{-\beta
r })^2}+\bigl[\bigl(2\kappa\nu\beta^2+\nu\beta^2\bigr)e^{-\beta r
}-\nu^2\beta^2e^{-2\beta r }\bigr]\,\frac{1}{(1-e^{-\beta r
})^2}\nonumber\\+\nu\beta^2\,\frac{e^{-\beta r}}{1-e^{-\beta r }}
+\beta\eta_{2}\left[E+M-A\right]\frac{e^{-\beta r}}{1-e^{-\beta r
}}+\left[E+M-A\right]\left[E-M\right]\bigg\}F(r)=0\,,\nonumber\\
\end{eqnarray}
where $\eta_{2}=\eta_{s}-\eta_{v}$. Using the same variable gives
\begin{eqnarray}
\left\{\frac{d^2}{dt^2}-\frac{t}{t(1-t)}\frac{d}{dt}+\frac{1}{t^2(1-t)^2}[-a^2_{1}-a^2_{2}t-a^2_{3}t^2]\right\}F(t)=0\,,
\end{eqnarray}
where
\begin{subequations}
\begin{align}
-a^2_{1}&=-\kappa(\kappa-1)+\nu(2\kappa-\nu+1)\,,\\
-a^2_{2}&=\nu(-2\kappa+2\nu)+\frac{\eta_{2}}{\beta}\,(E+M-A)\,,\\
-a^2_{3}&=-\nu(\nu+1)-\frac{\eta_{2}}{\beta}\,(E+M-A)+\frac{1}{\beta^2}\,(E+M-A)(E-M)\,.
\end{align}
\end{subequations}
Following the same procedure we obtain the energy eigenvalues for
the case of spin symmetry
\begin{eqnarray}
2\sqrt{\kappa(\kappa+1)-\frac{1}{\beta^2}\,(E+M-A)(E-M)\,}
+2\sqrt{\nu(\nu+1)+\frac{1}{\beta}(E+M-A)\bigl(\eta_{2}-\frac{E-M}{\beta}\bigr)\,}\nonumber\\
+\sqrt{(1+2\kappa)^2-4\nu(1-\nu+2\kappa)\,}-2n-1=0\,,
\end{eqnarray}
and the radial eigenfunctions for the upper component of the Dirac
eigenfunction as
\begin{eqnarray}
F(t)\sim t^{a_1}(1-t)^{(1+A)/2}\,P_n^{(A,\,\,2a_1)}\,(2t-1)\,,
\end{eqnarray}
which gives us the other radial component from Eq. (8a) as
\begin{eqnarray}
G(t) \sim
\frac{t^{a_1}(1-t)^{(1+A)/2}}{M-E+A}\bigg\{P_n^{(A,\,\,2a_1)}\,(2t-1)\bigg[\beta(1-t)
\big(\frac{a_{1}}{t}-\frac{1+A}{2(1-t)}+\frac{\kappa\beta}{ln t
}+\frac{\beta\nu}{lnt}(1-t)\big)\bigg]\nonumber\\+\beta(1-
t)(n+A+2a_{1}+1)P_{n-1}^{(1+A,\,\,1+2a_{1})}\,(2t-1)\bigg\}\,.
\end{eqnarray}
The numerical energy eigenvalues for different quantum number
pairs ($n,\kappa$) and different parameter values which makes it
possible to see the effect of the tensor interaction on the energy
eigenstates are showed in Table I (for $\eta_{2}=2.5$ a.u.,
$\beta=0.5$ a.u., $A=0$). We see that energy values are positive
if the Dirac equation has spin symmetry [25-27]. We observe that
the energy eigenvalues increase while the principal quantum number
increases or spin-orbit quantum number $\kappa$ decreases. It
could be interesting to study the case if we take the tensor
interaction as a Coulomb-like potential. Eq. (2) has the following
form for $\beta r \rightarrow 0$
\begin{eqnarray}
U(r) \sim \frac{\nu}{r}\,\bigl(1-\beta r+\ldots\bigr)\,,
\end{eqnarray}
which gives us an attractive Coulomb potential for $|\nu|<0$ and a
repulsive one for $|\nu|>0$ for the first-order approximation. In
this case, we obtain the energy eigenvalue equation for the case
of pseudospin symmetry
\begin{eqnarray}
2\sqrt{\kappa(\kappa-1)-\nu(2\kappa-1-\nu)-\frac{1}{\beta^2}\,[E^2-M^2-A(E+M)]\,}
+2\sqrt{\frac{1}{\beta}(M-E+A)\bigl(\eta_{1}+\frac{E+M}{\beta}\bigr)\,}\nonumber\\
+\sqrt{(1-2\kappa)^2-4\nu(2\kappa-1-\nu)\,}+2n+1=0\,.\nonumber\\
\end{eqnarray}
which is valid for attractive Coulomb potential while the
eigenvalue equation for the case of spin symmetry is written as
\begin{eqnarray}
2\sqrt{\kappa(\kappa+1)-\nu(2\kappa+1-\nu)-\frac{1}{\beta^2}\,[E^2-M^2-A(E+M)]\,}
+2\sqrt{\frac{1}{\beta}(M+E-A)\bigl(\eta_{2}+\frac{M-E}{\beta}\bigr)\,}\nonumber\\
+\sqrt{(1+2\kappa)^2-4\nu(2\kappa+1-\nu)\,}-2n-1=0\,.\nonumber\\
\end{eqnarray}
It should be noted that the terms including the tensor interaction
in Eq. (11) behave like a centrifugal barrier if one chooses the
tensor interaction as a Coulomb-like potential. So, we could
except that the number of bound states increase because of the
contributions coming from the tensor terms to centrifugal barrier.
We summarize the numerical results in Table II for the case where
if we take the tensor interaction as a Coulomb-like potential
given in Eq. (47).

\section{Conclusions}
We have studied the approximate bound state solutions of the Dirac
equation for the Yukawa potential for the cases where the Dirac
equation has pseudospin and spin symmetry, respectively, in the
existence of a tensor interaction having a Yukawa-like form. We
have obtained the energy eigenvalue equations and the related
two-component spinor wave functions with the help of
Nikiforov-Uvarov method. We have presented the numerical results
of the energy eigenvalues for the cases of pseudospin and spin
symmetry in Table I to see the effect of tensor interaction on
bound state energies and seen that this contribution is to create
much strongly bound states.

\section{Acknowledgments}
This research was partially supported by the Scientific and
Technical Research Council of Turkey

\newpage

\begin{table}
\begin{ruledtabular}
\caption{Energy eigenvalues in units of a.u. of the Yukawa
potential for the cases of pseudospin and spin symmetry ($M=1$).}
\begin{tabular}{@{}cccccccccccc@{}}
\multicolumn{12}{c}{pseudospin symmetry}\\ \cline{1-12} & & &
\multicolumn{3}{c}{$\nu=0$} & \multicolumn{3}{c}{$\nu=0.1$} &
\multicolumn{3}{c}{$\nu=1.0$}
\\ \cline{4-6} \cline{7-9} \cline{10-12}
$n$ & $\kappa$ & state &  & $E<0$ &  &  & $E<0$ &  & $E<0$ \\
1 & -1 & $1s_{1/2}$ &  & 1.22470 &  &  & 1.22461 &  & 1.13610 \\
  & -2 & $1p_{3/2}$ &  & 1.46034 &  &  & 1.45258 &  & 1.23104 \\
  & -3 & $1d_{5/2}$ &  & 1.60000 &  &  & 1.58996 &  & 1.29851 \\
  & -4 & $1f_{7/2}$ &  & 1.69025 &  &  & 1.67938 &  & 1.34844 \\
2 & -1 & $2s_{1/2}$ &  & 1.13610 &  &  & 1.12562 &  & 0.84186 \\
  & -2 & $2p_{3/2}$ &  & 1.23104 &  &  & 1.21723 &  & 0.76922 \\
  & -3 & $2d_{5/2}$ &  & 1.29851 &  &  & 1.28329 &  & 0.69221 \\
  & -4 & $2f_{7/2}$ &  & 1.34844 &  &  & 1.33250 &  & 0.59863 \\
\multicolumn{12}{c}{spin symmetry}\\ \cline{1-12} & & &
\multicolumn{3}{c}{$\nu=0$} & \multicolumn{3}{c}{$\nu=0.1$} &
\multicolumn{3}{c}{$\nu=1.0$}
\\ \cline{4-6} \cline{7-9} \cline{10-12}
$n$ & $\kappa$ & state &  & $E>0$ &  &  & $E>0$ &  & $E>0$ \\
0 & -2 & $0p_{3/2}$ &  & 0.49589 &  &  & 0.57269 &  & 0.97850 \\
  & -3 & $0d_{5/2}$ &  & 1.54413 &  &  & 1.55360 &  & 1.58112 \\
  & -4 & $0f_{7/2}$ &  & 1.92856 &  &  & 1.92117 &  & 1.86174 \\
  & -5 & $0g_{9/2}$ &  & 2.08474 &  &  & 2.07607 &  & 2.01207 \\
1 & -2 & $1p_{3/2}$ &  & --- &  &  & --- &  & 0.17028 \\
  & -3 & $1d_{5/2}$ &  & 1.21217 &  &  & 1.26027 &  & 1.48195 \\
  & -4 & $1f_{7/2}$ &  & 1.99593 &  &  & 1.99371 &  & 1.96986 \\
  & -5 & $1f_{9/2}$ &  & 2.19280 &  &  & 2.19280 &  & 2.17118 \\
\end{tabular}
\end{ruledtabular}
\end{table}

\newpage

\begin{table}
\begin{ruledtabular}
\caption{Energy eigenvalues in units of a.u. for the case where
tensor interaction chosen as a Coulomb-like potential
($\eta_{1}=\eta_{2}=2.5, M=5, \nu=0.5, \beta=0.5$).}
\begin{tabular}{@{}cccccccc@{}}
\multicolumn{4}{c}{pseudospin symmetry} & \multicolumn{4}{c}{spin
symmetry}
\\ \cline{1-4} \cline{5-8}
$n$ & $\kappa$ & state &  $E<0$ & $n$  & $\kappa$ & state & $E>0$ \\
1 & -1 & $1s_{1/2}$ &  2.86397 &  0 & -2 & $0p_{3/2}$ &--- \\
  & -2 & $1p_{3/2}$ &  3.36173 &    & -3 & $0d_{5/2}$ &0.86904\\
  & -3 & $1d_{5/2}$ &  3.57271 &    & -4 & $0f_{7/2}$ &2.18996\\
  & -4 & $1f_{7/2}$ &  3.67004 &    & -5 & $0g_{9/2}$ &2.78225\\
2 & -1 & $2s_{1/2}$ &  3.40594 &  1 & -2 & $1p_{3/2}$ &--- \\
  & -2 & $2p_{3/2}$ &  3.62177 &    & -3 & $1d_{5/2}$ &--- \\
  & -3 & $2d_{5/2}$ &  3.70932 &    & -4 & $1f_{7/2}$ &0.03410\\
  & -4 & $2f_{7/2}$ &  3.74219 &    & -5 & $1f_{9/2}$ &1.59973\\
\end{tabular}
\end{ruledtabular}
\end{table}


\newpage



\begin{thebibliography}{99}



\bibitem{ref1} H.~Yukawa, Proc. Phys. Math. Soc. Jpn. {\bf 17}, 48 (1935).





\bibitem{ref2} E.~R.~Vrscay, Phys. Rev. A {\bf 33}, 1433 (1986).





\bibitem{ref3} J.~P.~Gazeau and A.~Maquet, Phys. Rev. A {\bf 20}, 727 (1979).





\bibitem{ref4} S.~A.~Khrapak, A.~V.~Ivlev, G.~E.~Morfill and S.~K.~Zhdanov, Phys. Rev. Lett. {\bf 90}, 225002 (2003).





\bibitem{ref5} J.~C.~Inkson, \textit{Many-Body Theory of Solids} (Plenum, New York, 1986).



\bibitem{ref6} O.~A.~Gomes, H.~Chacham and J.~R.~Mohallem, Phys. Rev. A {\bf 50}, 228 (1994).



\bibitem{ref7} A.~Arda and R.~Sever, J. Math. Phys. {\bf 52}, 092101 (2011).



\bibitem{ref8} O.~Aydogdu and R.~Sever, Phys. Scr. {\bf 84}, 025005 (2011).



\bibitem{ref9} M.~R.~Setare and S.~Haidari, Phys. Scr. {\bf 81}, 065201 (2010).



\bibitem{ref10} E.~Z.~Liverts and V.~B.~Mandelzweig, Ann. Phys. {\bf 324}, 388 (2009).



\bibitem{ref11} B.~Gulveren, A.~Demirtas and R.~Ogul, Phys. Scr. {\bf 64}, 277 (2001).



\bibitem{ref12} S.~Panchanan, R.~R.~Choudhury and Y.~P.~Varshni, Can. J. Phys. {\bf 74}, 136 (1996).



\bibitem{ref13} S.~E.~Korenblit and Y.~V.~Parfenov, Physics of Atomic Nuclei {\bf 56}, 483 (1993).



\bibitem{ref14} E.~Papp, Phys. Scr. {\bf 43}, 14 (1991).



\bibitem{ref15} P.~Burt and W.~K.~R.~Watson, Nuovo Cimento {\bf 27}, 525 (1963).



\bibitem{ref16} G.~Moreno and A.~Zepeda, J. Phys. B {\bf 17}, 21 (1984).



\bibitem{ref17} T.~Imbo, A.~Pagnamenta and U.~Sukhatme, Phys. Lett. A {\bf 105}, 183 (1984).



\bibitem{ref18} J.~M.~Ugalde, C.~Sarasola and X.~Lopez, Phys. Rev. A {\bf 56}, 1642
(1997).



\bibitem{ref19} A.~D.~Alhaidari, H.~Bahlouli and M.~S.~Abdelmonem, J. Phys. A {\bf 41}, 032001 (2008).



\bibitem{ref20} B.~Gönül, K.~Köksal and E.~Bakir, Phys. Scr. {\bf 73}, 279 (2006).



\bibitem{ref21} M.~Grant and C.~S.~Lai, Phys. Rev. A {\bf 20}, 718 (1979).



\bibitem{ref22} F.~J.~Rogers, H.~C.~Graboske and D.~J.~Harwood, Phys. Rev. A {\bf 1},
1577 (1970).



\bibitem{ref23} K.~T.~Hecht and A.~Adler, Nucl. Phys. A {\bf 137}, 139 (1969).



\bibitem{ref24} A.~Arima, M.~Harvey and K.~Shimizu, Phys. Lett. B {\bf 30}, 517 (1969).



\bibitem{ref25} J.~N.~Ginocchio, Phys. Rev. Lett. {\bf 78}(3), 436 (1997).

\bibitem{ref26} P.~J.~Borycki, J.~Ginocchio, W.~Nazarewicz and M.~Stoitsov, Phys. Rev. C {\bf 68}, 014304 (2003).

\bibitem{ref27} A.~L.~Blokhin, C.~Bahri and J.~P.~Draayer, Phys. Rev. Lett. {\bf 74}, 4149 (1995).

\bibitem{ref28} C.~S.~Jia, P.~Guo and X.~L.~Peng, J. Phys. A {\bf 39}, 7737 (2009).



\bibitem{ref29} G.~F.~Wei and S.~H.~Dong, Phys. Lett. A {\bf 373}, 49 (2008).



\bibitem{ref30} Y.~Xu and S.~J.~Zhu, Nucl. Phys. A {\bf 768}, 161 (2006).



\bibitem{ref31} L.~H.~Zhang, X.~P.~Li and C.~S.~Jia, Phys. Scr. {\bf 80}, 035003 (2009).



\bibitem{ref32} G.~F.~Wei and S.~H.~Dong, Phys. Lett. A {\bf 373}, 2428 (2009).



\bibitem{ref33} O.~Aydogdu and R.~Sever, Eur. Phys. J. A {\bf 43}, 73 (2010).



\bibitem{ref34} R.~Lisboa, M.~Malheiro, A.~S.~de Castro, P.~Alberto and M.~Fiolhais, Phys. Rev. C {\bf 69},
024319 (2004).



\bibitem{ref35} W.~Greiner, Relativistic Quantum Mechanics (Springer Verlag, 1981).



\bibitem{ref36} A.~F.~Nikiforov, and V.~B.~Uvarov, \textit{Special Functions of
Mathematical Physics }, (Birkh\"{a}user, Basel, 1988).



\bibitem{ref37} R.~L.~Greene and C.~Aldrich, Phys. Rev. A {\bf 14}, 2363 (1976).



\bibitem{ref38} Y.~Xu, S.~He and C.~S.~Jia, J. Phys. A {\bf 41}, 255302 (2008).



\bibitem{ref39} L.~H.~Zhang, X.~P.~Li and C.~S.~Jia, Phys. Lett. A {\bf 372}, 2201 (2008).



\bibitem{ref40} M.~Abramowitz, and I.~A.~Stegun, (eds.),
\textit{Handbook of Mathematical Functions with Formulas, Graphs,
and Mathematical Tables} (New York, 1965).
\end{thebibliography}
\end{document}